\documentclass{mn2e}
\usepackage{graphicx}

\title{On the metallicity of the Milky Way thin disc and photometric abundance scales}
\author[M. Haywood]
       {M. Haywood\thanks{email : Misha.Haywood@obspm.fr} \\
        GEPI, Observatoire de  Paris, F-92195 Meudon Cedex,France}
\date{Accepted.
      Received ;
      in original form }
\pagerange{\pageref{firstpage}--\pageref{lastpage}}
\pubyear{2002}

\begin{document}         

\maketitle
\begin{abstract}

The mean metallicity of the Milky Way thin disc in the solar neighbourhood is still a matter of debate, and has 
recently been  subject to upward revision (Haywood, 2001). 
Our star sample was drawn from a set of solar neighbourhood dwarfs with photometric metallicities. 
In a recent study, Reid (2002) suggests that our metallicity calibration, based on Geneva photometry, is biased.
We show here that the effect detected by Reid is not a consequence of our adopted metallicity scale, and we confirm
that our findings are robust.
On the contrary, the application to Str\"omgren photometry  of the Schuster \& Nissen metallicity scale
is problematic. Systematic discrepancies of about 0.1 to 0.3 dex affect the photometric metallicity 
determination of metal rich stars, on the colour interval 0.22$< b-y <$ 0.59, i.e including F and G stars.
For F stars, it is shown that this is a consequence of a mismatch between the standard sequence $m_1(b-y)$ of the Hyades used
by Schuster \& Nissen to calibrate their metallicity scale, and the system 
of Olsen (1993, 1994ab).
It means that although Schuster \& Nissen calibration and Olsen photometry are intrinsically correct, 
they are mutually incompatible for metal rich, F-type stars.
For G stars, the discrepancy is most probably the continuation of the same problem, albeit worsened by the lack of 
spectroscopic calibrating stars.
A corrected calibration is proposed which renders the calibration of Schuster \& Nissen applicable to the catalogues of Olsen.
We also give a simpler calibration referenced to the Hyades sequence, valid over the same colour and metallicity ranges.
\end{abstract}
 
                    

\section{Introduction}

Solar neighbourhood stars serve as reference to which we compare the characteristics of the Galaxy
outside the immediate solar vicinity, and their properties scale our measurements of the galactic  structure and
evolution. In view of their importance, it is somehow surprising that their general properties, such as 
the mean metallicity of the galactic disc stars, is still a matter of debate. 
In a recent paper (Haywood 2001), we constructed a metallicity distribution from stars 
within 20pc from the sun.
This new metallicity distribution was shown to be centred on solar metallicity, 
or 0.1-0.2 dex higher than the value found in most previous studies.
We discussed that this discrepancy is the result of various biases that enter the definition of these
samples, the principal effect being caused by the selection of samples on the basis of
spectral type.
Another sensitive effect comes from the adopted metallicity scale.
The choice of a given metallicity scale is determined from two criteria :
the necessity to utilise  stars with as low masses as possible, in order to
avoid biases favouring young stars, and available photometry.
The most widely  used photometry  for studying the metallicity distribution is the
Str\"omgren photometry, with the metallicity scale from Schuster \& Nissen (1989) (hereafter SN).
In our own study, and  while the calibration of SN is given as valid down to $b-y$=0.59 (B-V$\approx$1.0), we 
used Geneva photometry and  the metallicity scale from Grenon (1978) for stars redder than
$b-y\approx0.42$ (B-V$\approx$0.67). The reason for this choice was that Geneva photometry is available
for a larger set of solar neighbourhood K dwarfs.
Hence, in the initial sample used by Haywood (2001), approximately half the stars
had their metallicity determined from Str\"omgren photometry, the other half from 
Geneva photometry.

In a recent paper, Reid (2002) finds that the (B-V,[Fe/H]) distribution of our sample 
shows a trend of about 0.2~dex from B-V=0.5 to B-V=1.0, 
suggesting  that the calibration from Grenon (1978) is plagued by a systematic error.
By constrast, his  (B-V, [Fe/H]) distribution, based entirely on the metallicity scale 
of SN, shows no such trend, if characterized by a simple linear regression.
Most recently, and while terminating our paper, a study on the same subject was 
presented by Twarog, Twarog \& Tanner (2002), exactly pointing to the problem we had 
discovered in Reid (2002) and that motivated the present work - a severe apparent deficiency in the metallicity scale of SN, causing 
underestimates of the photometric metallicity of late G type and early K type metal rich stars.

It appears however that the study by Twarog et al. has underestimated the 
discrepancy between photometric and spectroscopic abundances, in the sense that this
problem also affects F-type stars, in contrast with their claim.
Moreover, while they suggest that the origin of this discrepancy is an unsuspected high dependence of the metallicity on the $c_1$ index, 
we propose that it probably originates in a mismatch between the $m_1(b-y)$ standard sequence adopted in SN - which comes from Crawford (1975) - and the one that
can be deduced from the photometric catalogues of Olsen (1993, 1994ab) (see also Olsen (1984), table VI),
which constitute the vast majority of available measurements of Str\"omgren  photometry.
The aim of the present paper is threefold : 

(1) In section 2, we first present the discrepancy and quantify its amplitude on the whole colour range, that
is between  0.22$< b-y <$ 0.59.

(2) In section 3, we propose an explanation for the origin of this effect, and give a corrected calibration, for which
we calculate new coefficients on the basis of an enlarged spectroscopic dataset.
Using the same spectroscopic calibrating stars, we also give an alternative calibration using $\delta m_1$ and $\delta c_1$.

(3) Section 4 is a brief discussion of the impact of  the photometric calibration on the metallicity distribution of the solar neighbourhood stars.
In particular, it is demonstrated that the part of the sample in Haywood (2001) that is 
concerned by this effect is unchanged by the new calibration. 
That means that the results presented in Haywood (2001) are robust.

\begin{figure*}
\includegraphics[width=16cm]{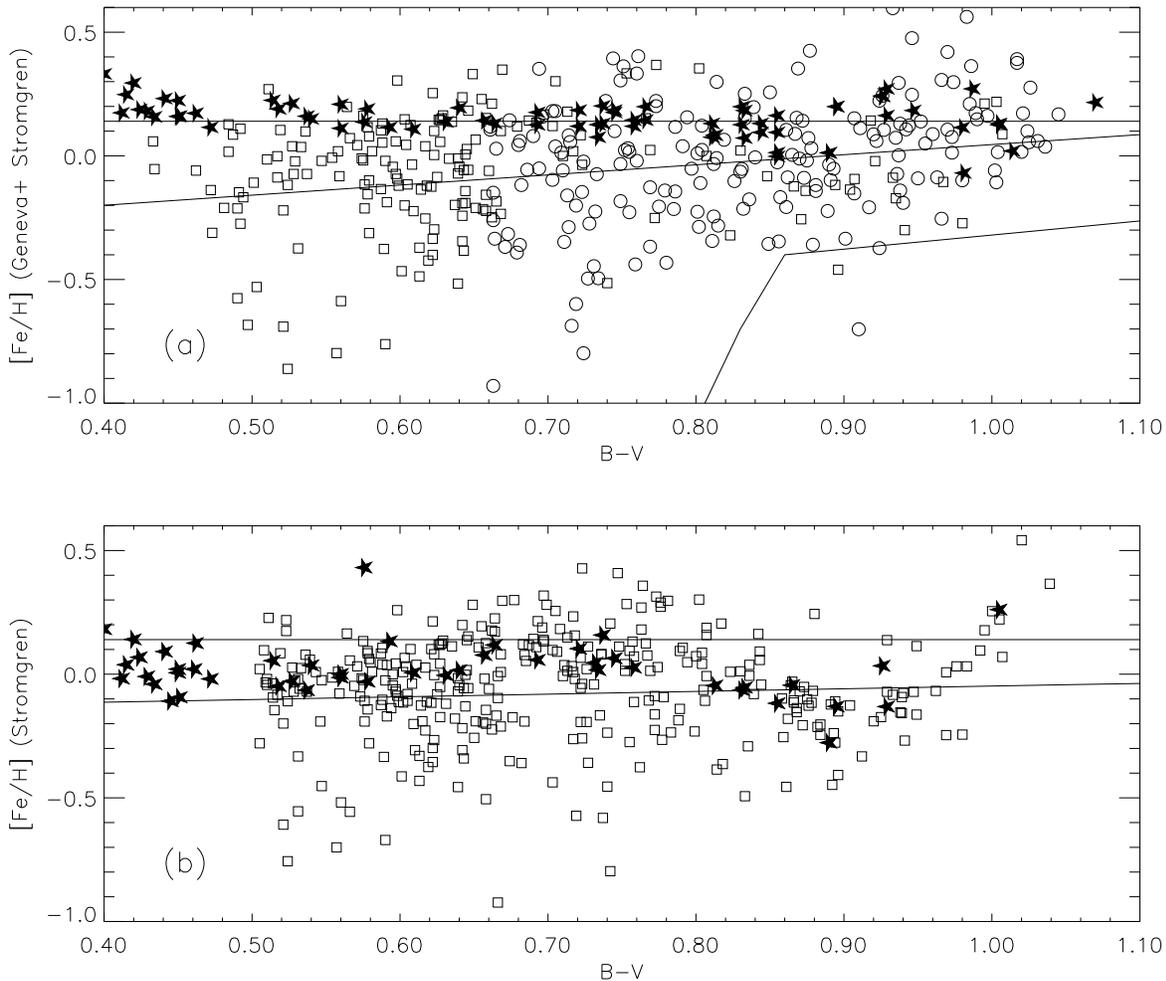}
\caption{Metallicity distribution as a function of 
B-V for our sample (a), and Reid (2002) sample (b). 
Our sample is a mixture of stars with Str\"omgren metallicity (squares) and Geneva metallicity (circles).
The horizontal line represents the metallicity of the Hyades cluster from Perryman et al. (1998), at [Fe/H]=+0.14.
The star symbols  are Hyades members from de Bruijne et al. (2001), with metallicity determined from 
the calibration of Grenon (1978).
The region delimited on the right lower part of the plot is the region where Favata Micela \& Sciortino (1997) found no
objects in the their sample. 
In the sample of Reid (2002) (plot b), all stars have their metallicity determined from Str\"omgren photometry, 
through Schuster \& Nissen (1989) calibration. Star symbols also represent Hyades members for which 
metallicity has been calculated through the calibration of  Schuster \& Nissen (1989). 
Our Str\"omgren metallicities in plot (a) were determined with the calibrations of SN but corrected as described in Haywood (2001), and
are not strictly equal to those of Reid (2002), although patterns common to the two ($B-V$,$[Fe/H]$) distributions can be 
seen.
}
\end{figure*}

\section{The colour-metallicity distribution}

Reid (2002) suggested that the Geneva calibration 
used in Haywood (2001) may be  affected by a colour term. This is illustrated in his Fig. 6b,
also shown  here (Fig. 1a) for convenience. The  linear regression on  
plot(a) shows that  there  is a trend of    metallicity with colour.   We  note that
although Reid (2002)  presents the trend  as being due to the Geneva
metallicity calibration,  his fit  is however made on the entirety of our sample,
which  is a mixed of Geneva  {\it and}  Str\"omgren metallicities  (the
Geneva   metallicities  being     used   for    stars   redder    than
B-V$\approx$0.67).   His  remark stems  from  the fact  that  
Str\"omgren  metallicities (using the calibration of SN) in
his sample  show no such trend  (regression line on Fig. 1b).  
It implies {\it a priori} that our  calibration for red stars is
questionable.   
Reid does not seem to envisage an intrinsic trend in our sample nor an effect caused by the metallicity scale of SN. 
He suggests that the trend is caused by the metallicity of the reddest stars in our sample.
However, looking  at  the   Fig. 6a in Reid (2002) (our Fig. 1b), we find the strangely distorted feature that
makes the upper part of the colour-metallicity distribution rather suspect 
(it seems to imply that there are no stars with super-solar metallicities at B-V=0.85-0.95).  
In order to elucidate the origin of these different trends, we decided to investigate in more 
detail the behavior of these two calibrations. 

There may  be three  reasons  for the  trend  reflected in the linear regression of Fig. 1a :

\begin{enumerate}
\item The Geneva calibration is biased,  as suggested by  Reid (2002), giving overestimated metallicities for red (B-V$>$0.8) objects.
\item The Str\"omgren calibration is biased, giving underestimated metallicities in the blue (B-V$<$0.67).
      But then, we must understand why Reid's (2002) sample shows no apparent trend in colour.
\item The trend is real and is determined by the sample selection. But then it also must be explained
why Reid's sample shows no such effect.
\end{enumerate}

There are two different ways we can check the metallicity calibrations. 
We can use cluster data, and test each calibration on clusters with well determined metallicities. 
The Hyades sequence is well appropriate for such work, and will be used below.
A second way is to use spectroscopic data. A number of spectroscopic
metallicities have been published in the recent years. And while they are still relatively sparse for K type
stars, we shall pay particular attention to possible colour effects.
To complement these two, we can search for systematic effects with the position of stars in the HR diagram. 
For example, 
the scarcity of stars redder than B-V=0.8 and more metal-poor than [Fe/H]=-0.4 contributes to the trend 
that is seen on Fig. 1a (we note, however, that this feature is also present in the Str\"omgren sample, Fig. 1b). 
 We may check if a corresponding feature is also seen in the HR diagram in our sample.

\begin{figure}
\includegraphics[width=8cm]{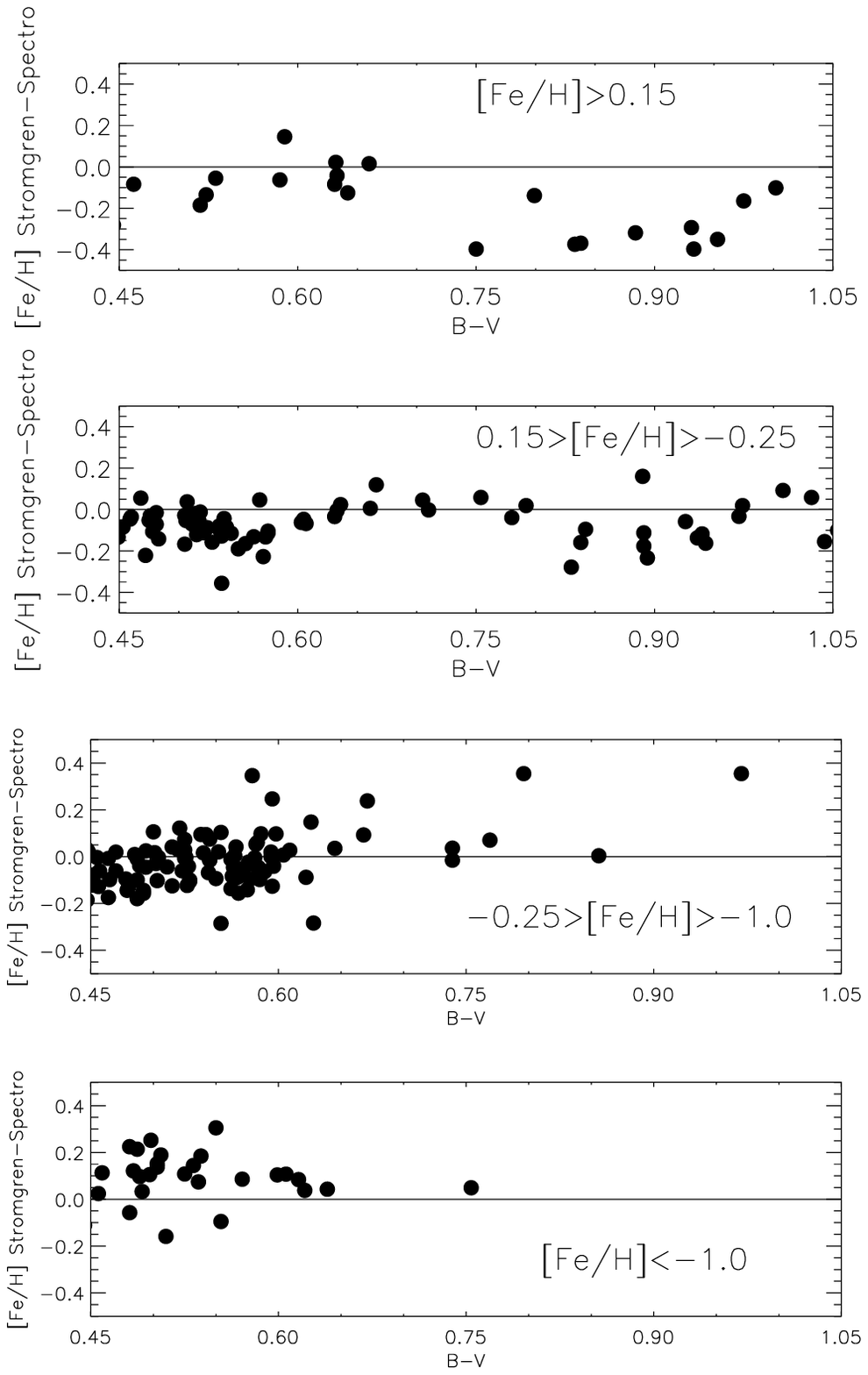}
\caption{Difference between photometric (Str\"omgren) and spectroscopic
metallicities, for 3 metallicity intervals. 
}
\end{figure}
\begin{figure}
\includegraphics[width=8cm]{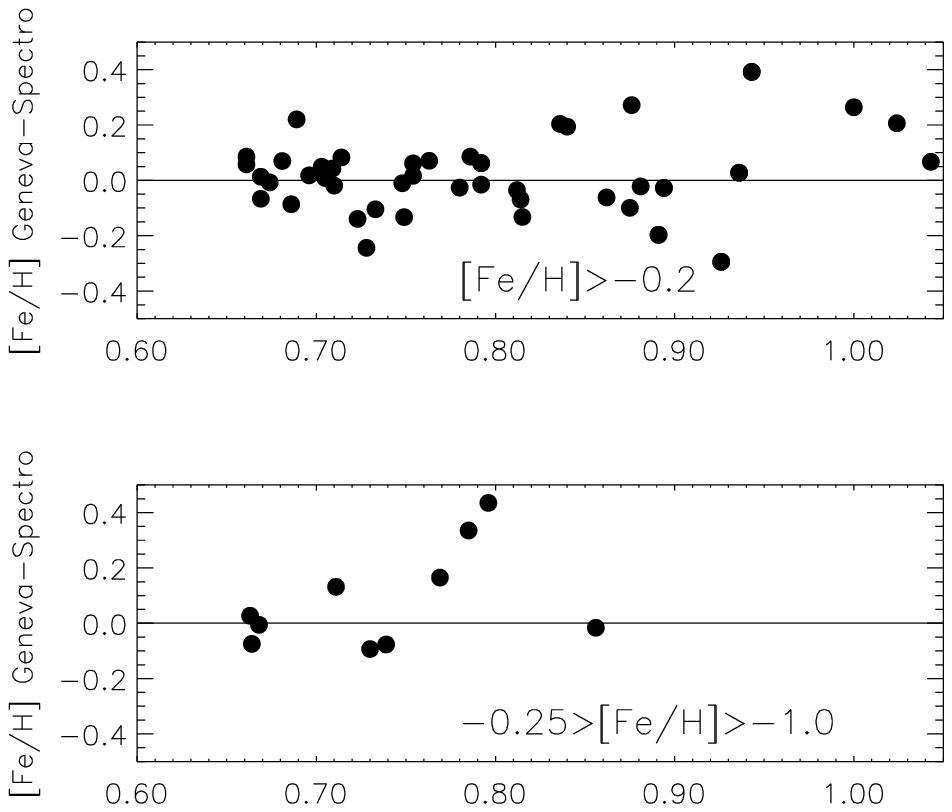}
\caption{Difference between photometric (Grenon (1978)) and spectroscopic
metallicities, for 2 metallicity intervals. 
}
\end{figure}

\subsection{The Hyades cluster test}

Fig. 1ab  shows the (B-V,  [Fe/H]$_{photo}$)  diagrams for the
samples of Haywood (2001)  and Reid (2002).   The horizontal line shows
the median metallicity  of  the Hyades cluster  as  determined from  a
sample   of spectroscopic metallicities in  Perryman   et al. (1998) at
[Fe/H]=+0.14.    The star   symbols show  Hyades members  with
metallicity  as determined from the calibration  of Grenon (Fig. 1a) and
the calibration of SN (Fig. 1b).  Since  the calibration by Grenon (1978)
was designed  by reference to the  Hyades sequence, it is not surprising (but reassuring)
that the plot  shows consistency with the  spectroscopic determination.

Different is the result of  the calibration of SN applied to the
Hyades  cluster members,  as  seen  on  Fig. 1b.   The   figure shows  
systematic deviations from the Hyades metallicity, which follow those of field stars.  
The Hyades metallicity is underestimated  by   about 0.15  dex  at
B-V=0.5-0.60, and  the  effect is even  stronger  at B-V$>$0.8 with  a
0.3-0.4~dex offset. This last feature is the one discussed by Twarog et al. (2002). 
In the blue part of the interval, it is not clear
how  much solar metallicities  (i.e  [Fe/H]$\approx$0)
would be affected by the bias, at least solely from the Hyades
cluster test.  The bias may affect  only metal-rich stars.   At redder
colours, it is however probable that even solar metallicity stars have
their  metallicities underestimated by  at least 0.15~dex. This can be
checked on spectroscopic data (section 2.2). 
While  we postpone our explanation to section 3, it is worth  
noting here that the calibration itself  is not responsible for the effect that
is seen on Fig. 1b at $b-y<0.375$. It is its {\it application} to available Str\"omgren photometry which is in question. 
What is relevant at the moment is that 
the (B-V,[Fe/H]$_{SN}$) distribution of Fig. 1b is strongly biased and the 
apparent lack of correlation given by the linear regression is meaningless.

Reid (2002) notes a general offset of about 0.1~dex in the Str\"omgren metallicity scale, 
following the similar result of  Alonso, Arribas \& Martinez-Roger (1996) and Haywood (2001).
Since all the comparisons in Reid (2002) are relative and made in the system of the SN calibrations, the zero point 
is {\it a priori} not crucial. 
However, the differential effect in colour is sufficiently strong that
Reid (2002) acknowledges that  his metallicity distribution (after rescaling by 0.1~dex) still has a less
extended tail towards metal   rich   stars  compared to   Favata  et
al. (1997) and  Haywood (2001). Although Reid (2002) does not  question further on the possible
origin of this discrepancy, the direct reason  for  this effect  is 
particularly conspicious  on the (B-V,[Fe/H]$_{SN}$)  distribution:
SN metallicities  are artificially lowered  below
[Fe/H]=-0.1  at 0.8$<$B-V$<$0.95  and below [Fe/H]=0.0  at B-V$<$0.65.

\subsection{Spectroscopic check}

\subsubsection{Schuster \& Nissen (1989)}

A second check can be done using  spectroscopic metallicities from the
literature.   We have used mainly data   from Edvardsson et al. (1993),
Favata et al (1997), Fulbright (2000), Feltzing \& Gustafsson (1998),
plus some additional  objects  from the catalogue of  metallicities by Cayrel de Strobel, Soubiran \&  Ralite  (2001)
for red (B-V$>$0.65)  objects.  Fig. 2 shows the
difference    between  photometric   (SN)  and  spectroscopic
metallicities,  for    4    metallicity  intervals.    Around   solar
metallicities, the colour variation is  similar to the one detected on
the Hyades.  It  confirms    that the   calibration of SN underestimates
metallicities  above [Fe/H]$>$-0.25 by 0.1-0.2~dex for stars bluer than  B-V$<$0.65, and by
0.1-0.3~dex  in the colour interval  0.75$<$B-V$<$0.9.   At    -0.25$>$[Fe/H]$>$-1.0,  the
combination of the 2   distinct  calibrations is apparent,   with  the
calibration for $b-y>$0.375 overestimating  the metallicity, and the calibration at  $b-y<$0.375 underestimating
metallicities.  
Finally, the  calibration    seems   to    overestimate  slightly the   metallicity   at [Fe/H]$<$-1.0. 

Because the calibration of SN has been applied mostly to F stars, 
the discrepancy of about 0.1~dex that is visible at B-V$<$0.65 and for [Fe/H]$>$-1.0 has already been 
noted by different authors, see in particular Alonso  et al. (1996).
In our own sample, it has been corrected as [Fe/H]=[Fe/H]$_{SN}$/0.86+0.05 (Haywood , 2001).

\subsubsection{Grenon (1978)}

In Haywood (2001), we checked the calibration of Grenon (1978) with spectroscopic determinations. 
Fig. 3 shows the result of this calibration compared to the same data as for the calibration of SN,
within colour limits that define  the calibration, that is 0.40$<B_2-V_1<$0.65. 
The data is very sparse for red metal poor dwarfs, and not much can be said about
the calibration below [Fe/H]$\approx$-0.25 dex. At [Fe/H]$>$-0.2,  the spectroscopic data
shows no offset. It it clear however that a secure calibration would require a larger data set.

The consistency of the Geneva metallicity scale can also be checked on the HR diagram.
The Hyades  cluster  was  already used  in Haywood (2001). We  further   detail our 
comparison with  the  cleaner Hyades sequence of de Bruijne et al (2001).  Fig. 4 
shows  our sample with the Hyades sequence.

\begin{figure}
\includegraphics[width=8.5cm]{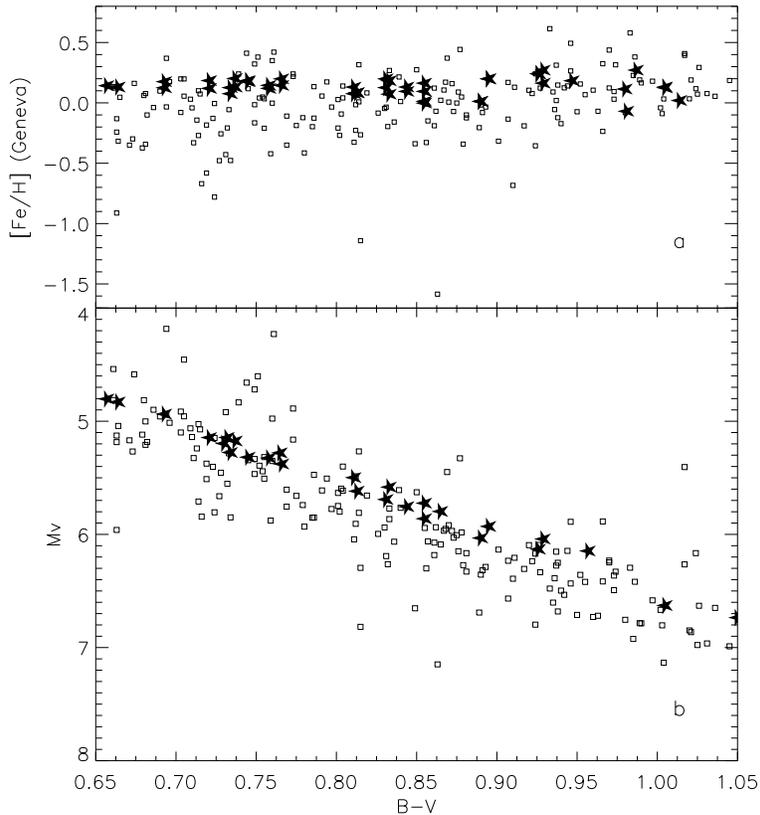}
\vspace{0.5cm}
\caption{The (B-V,metallicity) distribution of the stars in our sample with Geneva metallicities,
and the corresponding HR diagram. Star symbols are the Hyades, selected by
deBruijne et al. (2001). There is a good overall consistency between the metallicity of 
the stars and their position in the HR diagram. See text for details.
}
\end{figure}

\noindent We note 2 features :
\begin{enumerate}
\item There are no stars above the Hyades sequence around  $B-V$=0.9 in 
metallicity. A corresponding feature is visible in the HR diagram, Fig. 4b.
Only three stars stand clearly above the Hyades sequence between 
0.82$<B-V<$0.93, consistently in both diagrams.
\item There are no field stars in the HR diagram at  B-V$>$0.87 and M$_v >$ 7.2. 
Such  dwarfs would be expected to have [Fe/H]$<$-0.5, and we note a corresponding lack of stars 
in Fig. 1a (one star only). It is appropriate to remind that this feature
is visible on many studies  of the solar neighbourhood metallicity distribution 
(even with pre-Hipparcos parallaxes). The first study to mention this effect was
Favata et al. (1997) (see also Flynn \& Morell (1997), their Fig. 5). 
The samples considered in solar neighbourhood studies are too small to 
investigate whether this is only a statistical effect or a real absence
of midly-deficient stars at that colour.
\end{enumerate}
In the HR diagram at $B-V<$0.7, field stars above the Hyades sequence probably have evolved off the ZAMS
(the turn-off $B-V$ colour for solar metallicity stars at 12~Gyrs is circa 0.70).
At the bottom of the HR diagram (B-V$>$ 0.95), 4 stars with [Fe/H]$>$0.14 lie clearly below 
the Hyades sequence, which may be the signature of a bias overestimating metallicities in the Grenon calibration
at B-V$>$ 0.95. Among these 4 stars, only HIP 116745 has a measured spectroscopic metallicity
at [Fe/H]=-0.22 in the catalogue of metallicities of Cayrel de Strobel et al. (2001).

We conclude that there is no significant deviation between the HR diagram and
the colour-metallicity distribution of Fig. 1a.

\begin{figure}
\includegraphics[width=9cm]{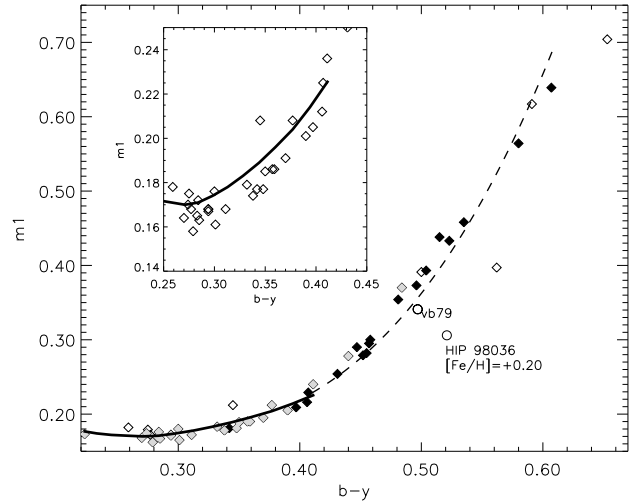}
\caption{The Hyades sequence as given by the Str\"omgren photometry from the GCPD (diamond). 
The $b-y$ and $m_1$ indices that are used are (in most cases) the mean of several measurements as given by the GCPD. 
In case the only source of photometry is  Olsen (1993) and/or Olsen (1994ab), the symbol
is a black filled diamond. 
If the photometry is the mean of Olsen and other sources, the symbol is a grey-filled diamond.
If the photometry contains no measurement from either   Olsen (1993) or Olsen (1994ab), the symbol is an empty diamond.
The continuous line is the standard sequence of Crawford (1975), which is the main data used by SN to calibrate their
metallicity scale at 0.22$<b-y<$0.375 for metallicities around 0.13~dex. The offset between the Crawford sequence and
the Hyades from the GCPD is of the order of 0.01 mag. 
For redder stars, it is suggested that, in practice, the calibration by SN follows 
an extrapolated sequence ressembling the one plotted as a dashed line.
It is interesting to note that the 2 outliers at $b-y$=0.345 and $b-y$=0.259, are stars which have not been observed by Olsen.
The open circles are the only  2 stars we could find with both Str\"omgren photometry at the time SN established their calibration,
and metallicity above solar. One of them is the Hyades  star vb79.
}
\end{figure}

\section{The Str\"omgren metallicity scale}

The calibration designed by Schuster \& Nissen (1989) has proved 
to be most useful for a variety of studies using Str\"omgren photometry,
and this has become still more evident with the advent of Hipparcos data.
Results of Sec. 2 shows however that its application to available data is problematic
for metal rich stars. We now try to explain the probable cause of that problem.

\subsection{A diagnostic}

Contrary to the claim of Twarog et al. (2002),  the Str\"omgren metallicities at 0.22$<b-y<$0.40-0.41 (or 0.35$<$B-V$<$0.65) are also 
affected, with the Hyades having their abundance underestimated by 0.1-0.2~dex. Such strong effect is somewhat puzzling since 
the F-star calibration of SN is said to be constrained using the standard (Hyades) sequence of Crawford (1975) at a metallicity of 0.13~dex
(and it is given double weight in the calibration of SN).
This failure is interesting because it gives us some clue on the possible general  origin of the problem.
Since the photometric Hyades metallicity adopted  by SN should be consistent with the spectroscopic value at [Fe/H]$\approx$ 0.13, 
the only possible cause for the defect that is seen on Fig. 1 must come from the Str\"omgren indices that are used to 
represent the Hyades sequence.

\begin{enumerate}
\item F stars 0.22$<b-y<$0.375 \\
Fig. 5 shows the $m_1(b-y)$ sequence for the Hyades, on the relevant colour range, from two different sources.
The continuous line is the standard sequence from  Crawford (1975) used by SN to calibrate their metallicities.
Diamond symbols are Hyades members with Str\"omgren photometry  from the 
GCPD (The General Catalogue of Photometric Data, Mermilliod, Mermilliod, \& Hauck, 1997). 
The $b-y$ and $m_1$ indices of Fig. 5 are the mean of measurements from different sources as given by the GCPD.
Most of these points include one or several  measurements  by  Olsen (1993) and/or  Olsen (1994ab).
These various origins  have been differentiated on Fig. 5 as follows :
Black diamonds are those stars for which  indices contain only measurement 
from Olsen (1993) and/or  Olsen (1994ab). Grey diamonds  are a mixed of Olsen (1993 and/or 1994ab)
and measurements from other sources.
Finally, empty diamonds represent those stars which were not observed by  Olsen (1993 and 1994ab).
Since the Hyades sequence is mostly dominated by the measurements of Olsen, we call it the `Olsen sequence' 
herebelow.  
In the $b-y$ range of interest ($b-y<$0.375), the two $m_1(b-y)$ scales (Crawford (1975) and the  Olsen sequence) 
clearly show an offset of about $\Delta m_1 \approx$0.01 mag.
For F-type stars, $m_1$ is the index sensitive to the metallicity, with  $\Delta m_1$=0.01 mag
corresponding to  $\Delta$[Fe/H]$\approx$0.1dex  on the metallicity scale.

Since the huge database established by Olsen is the main source for Str\"omgren photometry, and since
the calibration of SN is based on the sequence of Crawford (1975), it is expected that metallicity estimates 
that combine both will be incorrect by about 0.1~dex, even though the calibration of SN is correct and Olsen indices are 
precise.\\

\item G stars 0.375$<b-y<$0.59

The continuity in the deviations of metallicity estimates on Fig. 1b accross the limit that separate the 2 calibrations suggest
a similar origin for the defects at $b-y<0.375$, and $b-y>0.375$.
It is seen on Fig. 1b that the photometric metallicity offset of 0.1~dex is continuous accross the two intervals, 
i.e it is left unaffected by the change of calibration, at least up to $b-y$=0.40.
This comes in support to  the fact that the discrepancy  originates in the  indices
rather than in the calibration themselves.

At 0.375$<b-y<$0.59, and for the metal rich part of their calibration, SN have used the 2 reddest points
in the sequence of  Crawford (1975) (at $b-y$=0.394 and 0.412), and  `to give more weight to the Hyades, 
four individual stars with an average [Fe/H]=+0.09 from Cayrel
de Strobel et al. (1985) and from Cayrel et al. (1985) were included with single weight'.
We could find  only one star from the Hyades satisfying this description and that had Str\"omgren indices
measurement available in 1989. This star (vb79), sufficiently red to have been useful for calibrating the
reddest part of the metallicity scale, has $b-y$=0.497 and $m_1$=0.341 (Carney, 1983).
Another star from the catalogue of  Cayrel de Strobel et al. (1985), not belonging to the Hyades
cluster, but with [Fe/H]$_{spectro}$=0.25 is also shown on Fig. 5.
 
On Fig. 5, we have extrapolated the standard sequence of Crawford (1975) at $b-y>$0.41
with a polynomial. The calibration given by SN is not explicitly referenced to the Hyades 
sequence, but the standard sequence  is implicitly integrated in their functional form.
Although we can't quantify the difference between the Olsen sequence
and the one {\it de facto} adopted by SN,  we suggest
that the lack of data and the functional form of the calibration of SN has probably led to a sequence
resembling the one proposed as an extrapolation on Fig. 5.
Several features support this suggestion :\\

First, it is significant that the Crawford and extrapolated sequence on Fig. 5 crosses the Olsen sequence 
between 0.40$<b-y<$ 0.45 (0.65$<B-V<$0.74). It implies that on this interval,
both $m_1$ indices from Olsen or the extrapolated sequence overlap, and will give similar metallicities. 
As a matter of fact, this is precisely the interval where
the calibration of SN (using Olsen Str\"omgren indices), gives a metallicity for the Hyades nearest to the spectroscopic value.

Second, between 0.45 $<b-y<$ 0.57, the offset $\Delta m_1$ between the polynomial sequence and the Olsen sequence 
reaches 0.03-0.05, which precisely is the offset that is necessary in order to level up the photometric 
Hyades metallicities to 0.14~dex, if Olsen $m_1$ indices are used in the calibration of SN.

Third, the polynomial sequence rises more steeply than the observed sequence, which flattens 
to reach a maximum in $m_1$ at $b-y \approx$0.75.
It is expected that the extrapolated polynomial sequence rejoins the observed (Olsen) sequence at some
colour in the interval  $0.5<b-y<0.7$. That means that at this
colour, the two sequences should again give the same metallicity. This is precisely what is 
observed on Fig. 1 at $b-y=0.57$ (i.e the  calibration of SN and Olsen indices give [Fe/H]$\approx$+0.14, which is the 
metallicity expected from the standard relation).

\end{enumerate}

\begin{figure}
\includegraphics[width=9cm]{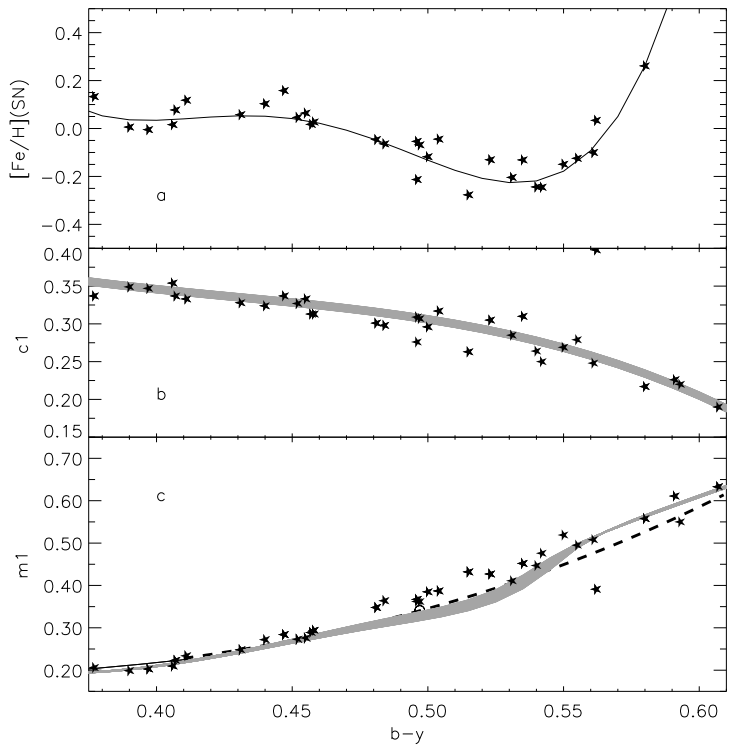}
\caption{
The polynomial Hyades sequence ($b-y$,$m_1$) (continuous line, plot c), as reconstructed from the ($b-y$,$[Fe/H]$),  ($b-y$,$c_1$) relations (plot a and b) 
and equation  (1). The broken line is the extrapolated sequence of Fig. 5.
}
\end{figure}

Altogether, these clues convey the impression that it is the offset between the two $m_1(b-y)$ sequences that
is responsible for the strong colour dependence and the systematic discrepancy seen on Fig. 1.
It is meaningful that the Crawford and extrapolated polynomial sequence of Fig. 5 lies respectively {\it above} and {\it below}
than the Olsen sequence in just the correct intervals to explain the behavior of the photometric
metallicity seen on Fig. 1b.

\subsection{Consistency  argument}

\begin{figure*}
\includegraphics[width=15cm]{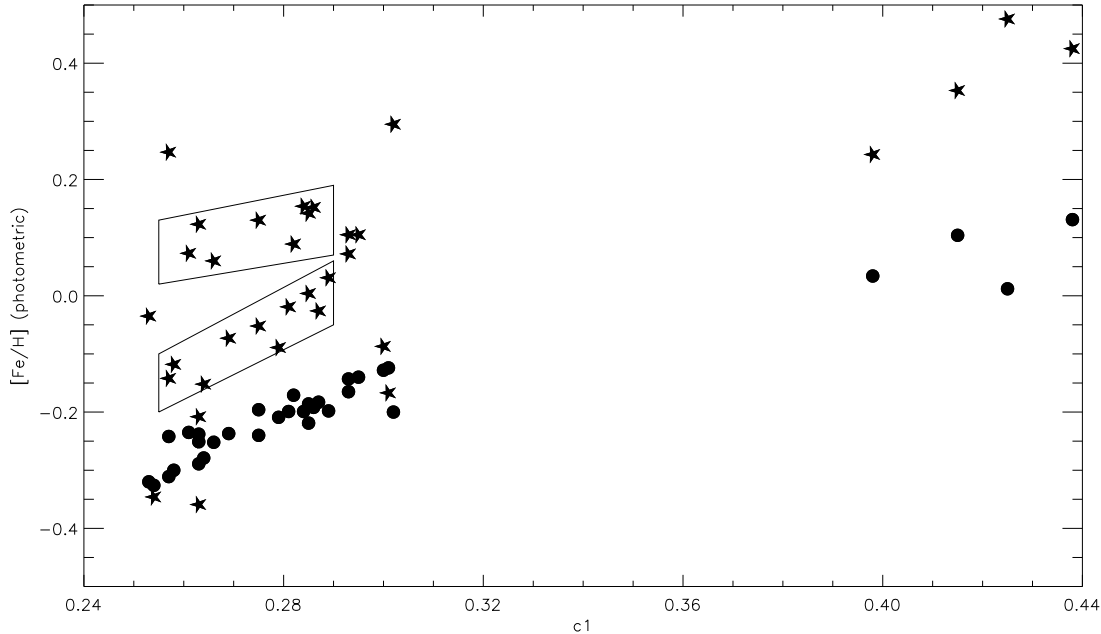}
\caption{$c_1$ dependence of photometric metallicities. Star symbols are metallicities from Geneva photometry,
dots are metallicities of SN, for the same objects. Objects in the upper box have a mean $m_1$=0.445, and $m_1$=0.408 in the 
lower box. See Sec. 3.3 for comments.
}
\end{figure*}

Following this last remark, it is possible to use the calibration given 
by SN to recover the Hyades sequence and check the consistency of our argument.
Assuming relations between ($b-y$, $c_1$) and  ($b-y$, $[Fe/H]_{SN}$), $m_1$ can be calculated
as the root of the  equation :

\begin{eqnarray}
(-53.8 + 145.5*b-y -137.2*c_1) * m_1^2    +  \nonumber\\  ( 22.45 + 85.1*c_1 - 62.04*b-y  ) * m_1 \nonumber\\  - 2.0965 -13*c_1^2 -[Fe/H]_{SN} =0
\end{eqnarray}

The ($b-y$,[Fe/H]$_{SN}$),  ($b-y$,$c_1$) relations for the Hyades have been assumed as polynomial fits to
the Hyades data, as shown on Fig. 6ab. Because we don't know the exact ($b-y, c_1$) sequence endorsed by 
SN at $b-y>0.4$,  we have assumed a set of ($b-y, c_1$) relations, as given on Fig. 6b.

They are used to calculated the coefficients of the polynomial (1), which
is solved for $m_1$, at different $b-y$.
The result is shown on Fig. 6c and is consistent with our discussion above.

\subsection{Twarog et al. (2002)}

In their paper, Twarog et al. (2002) proposed another cause for the discrepancy.
They state that the calibration of SN  underestimates the $c_1$ dependence of metallicity
for stars redder than b-y=0.47.
Their argument relies on their Fig. 5, which shows the correlation between metallicity (both spectroscopic 
and photometric) and $c_1$  for stars with b-y$>$0.47. 
If the argument is correct, we would expect
however that $c_1$ correlates differently for spectroscopic and 
photometric (SN) metallicities.
Their Fig. 5 seems to illustrate, at variance with their claim, that the correlation
is approximately similar for both spectroscopic and photometric abundances, which means 
that the photometric metallicities of SN are correctly tied to the $c_1$ index.

The interpretation that Twarog et al. give of their  Fig. 5 is qualitative
and relies on a dozen of stars. Among these, two objects that are outliers 
can give the favourable impression that the slopes of the photometric 
and spectroscopic [Fe/H]-$c_1$ data sets are different. 
Even if this was the case, it would remain to be demonstrated that this is
due solely to the $c_1$ index.

In order to extend the comparison of Twarog et al. (2002) to a greater number of stars, 
we use our Geneva photometric metallicities instead of the spectroscopic metallicity
scale. We select in our sample all stars having $0.85<B-V<0.95$ (equivalent to $0.50<b-y<0.55$),
giving 37 objects (32 are within the limits of Fig. 7).
We plot on Fig. 7 the photometric metallicities [Fe/H]$_{SN}$ and [Fe/H]$_{GEN}$ as a function of  $c_1$, a figure similar 
to Fig. 5 in  Twarog et al. (2002),
 but now with the Geneva photometric scale being the reference.
Of course, this  comparison relies on the correctness of Geneva metallicities.
Following the arguments developed in Sec. 2, we assume that they are indeed valid. 
Interestingly, it seems that our Fig. 7 reveals the patterns that were only suggested in the  Fig. 5 of Twarog et al. (2002).
That is, Geneva metallicities show that there is a 
second group of stars (upper box) standing 0.15-0.20 dex above the sequence at $c_1$=0.25-0.30.
Only a small part of that group is visible in Twarog et al. (2002), at $c_1$=0.28-0.30.

\begin{figure*}
\includegraphics[width=16cm]{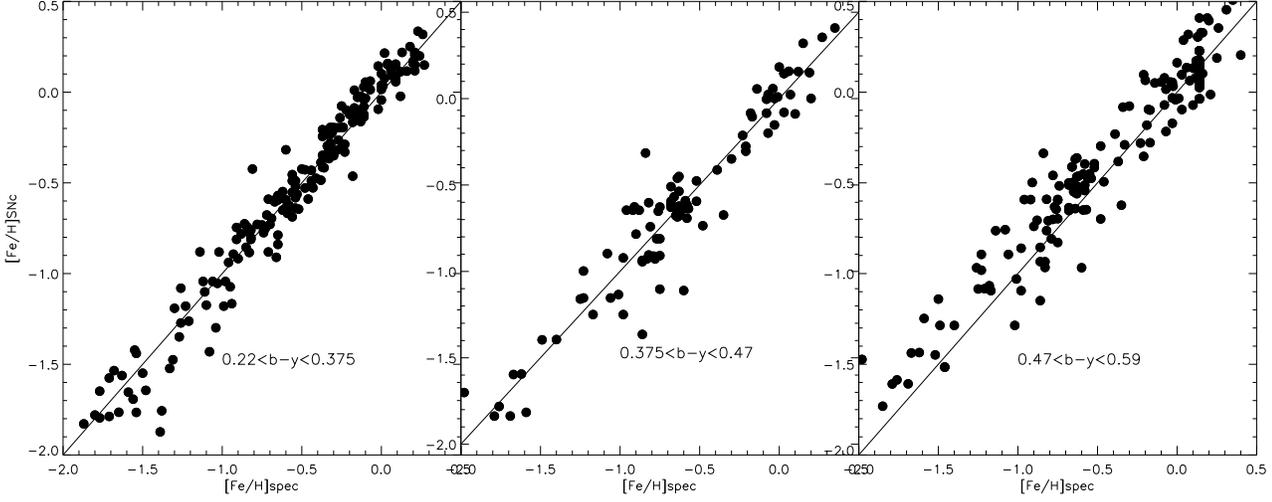}
\caption{Corrected photometric metallicities (eq. 2, 3, 4) $versus$ spectroscopic metallicities.
}
\end{figure*}

What is interesting  is that this group is not differentiated in the metallicities of SN.
That means the calibration of SN lacks a dependence  on one parameter, which clearly cannot 
be $c_1$. 
The mean and dispersion of $m_1$  for the eight stars in the upper box are 0.445 and 0.036.
The lower box contains 9 stars with a  mean $m_1$=0.408 and dispersion 0.030.
We can estimate  the corresponding difference in metallicity using Olsen (1984). 
The calibration of Olsen (1984) is valid for stars with $b-y \geq 0.514$, and the 
metallicity is proportional to $\delta m_1$ with a  coefficient equal to 5.1. Using this
coefficient, the difference in $m_1$ of our two  groups corresponds to 0.19~dex in metallicity,
similar to the difference given by  the Geneva photometry.

If we  restrict our comparison to stars in the lower box  only and those at
 $c_1>0.39$, it is true that the difference between the reference (Geneva) metallicities and SN metallicities
are  larger at  $c_1>0.39$. However, this could be only a statistical effect 
due to the lack of data at   $c_1>0.39$. In any case, the 4 stars at  $c_1>0.39$ are
a minority, and if we restrict our discussion to where the majority of the stars are 
(at  $c_1<0.31$), the idea of a difference of slope between the two metallicity  scales
is meaningless. 
We conclude that Fig. 7 confirms our argument, with $m_1$ being the source of 
the problem in the calibration of SN.

\subsection{A corrected SN calibration}

Taking into account the above remarks, we give a modified version of the Schuster \& Nissen
calibrations, based on newer calibrating metallicities. We keep the same functional
form as SN, but we redetermined the coefficients by a least square fitting procedure.
In order to find the best fitting coefficients, we use $PIKAIA$, a genetic algorithm developped
for optimization problems by P. Charbonneau (see Charbonneau, 1995), to which the reader should
refer for an extensive description of this technique. A `fitness' function must be defined for PIKAIA to optimize
the fit between a set of spectroscopic metallicities and the photometric metallicities.
Our fitness function is a least square minimization :\\

\begin{displaymath}
 \chi^2  = \sum_{i=1}^{N} [y(x_i,a_j)-y_i]^2,
\end{displaymath}

where  $y_i$ are the spectroscopic metallicities, and $y(x_i,a_j)$ are the photometric metallicities,
which are function of the photometric indices $x_i$ and coefficients $a_j$.\\

(a) {$0.22<b-y<0.37$ :}\\

We first run the optimization procedure on a set of spectroscopic metallicities comprising 
the data of Edvardsson et al. (1993), dwarfs in the sample of Fulbright (2001), and a few 
Hyades stars, to which we attribute a metallicity of [Fe/H]=+0.14. That amounts to 211 stars.
As in SN, the Hyades stars have been attributed a weight of 2.
The fit is valid over $-2.0<[Fe/H]<0.5$,

\begin{eqnarray}
[Fe/H]=-2.0-43.90m_1+353.4(b-y)m_1+\nonumber\\
18.0(b-y)m_1^2-612.6(b-y)^2m_1+(6\nonumber\\
-48m_1-7.85(b-y))(log(m_1-c_3))
\end{eqnarray}
 
and  $c3=0.627-7.04(b-y)+11.25(b-y)^2$\\

(b) {$0.37<b-y<0.47$ :}\\

The fit was made with 103 spectroscopic metallicities  and 13 Hyades stars at [Fe/H]=+0.14.

\begin{eqnarray}
[Fe/H]=-1.64 + 11.09m_1 - 29.29m_1^2 - 57.40(b-y)m_1 \nonumber\\ + 116.96m_1^2(b-y)+(128m_1-22.231c_1-206.48m_1^2)*c_1
\end{eqnarray}

(c) {$0.47<b-y<0.59$ :}\\

The fit was made with 36 spectroscopic metallicities and 12 Hyades stars at [Fe/H]=+0.14.
\begin{eqnarray}
[Fe/H]=-1.64 + 16.75m_1 - 12.61m_1^2 -52.17(b-y)m_1\nonumber\\ + 66.02 6m_1^2(b-y)+(47.98m_1-3.99c_1-65.06m_1^2)*c1
\end{eqnarray}

\begin{figure*}
\includegraphics[width=16cm]{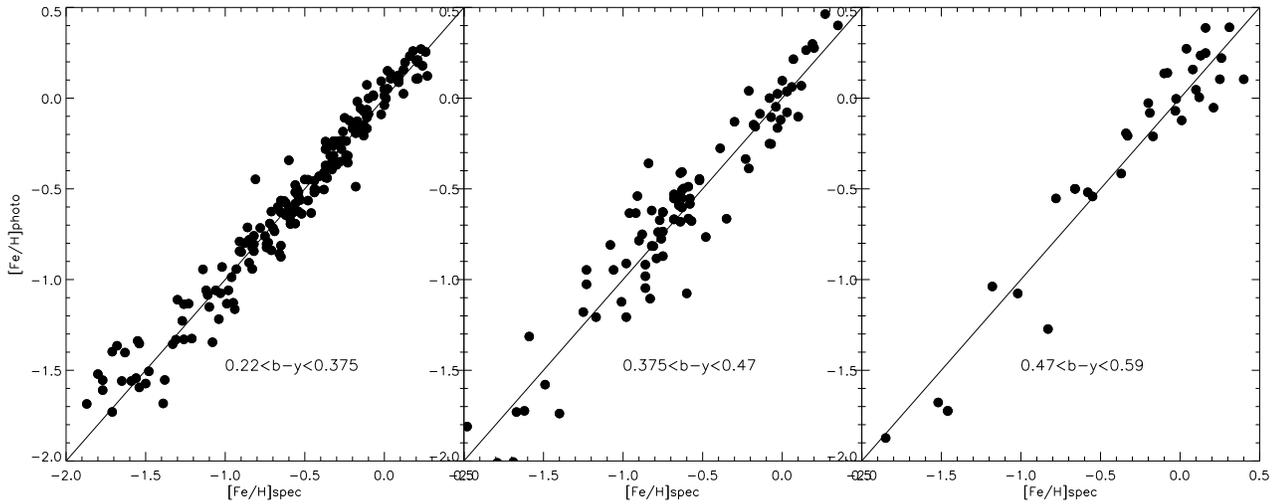}
\caption{Photometric metallicities (eq. 5 and 6) $vs$ spectroscopic metallicities.
}
\end{figure*}

\begin{figure*}
\includegraphics[width=16cm]{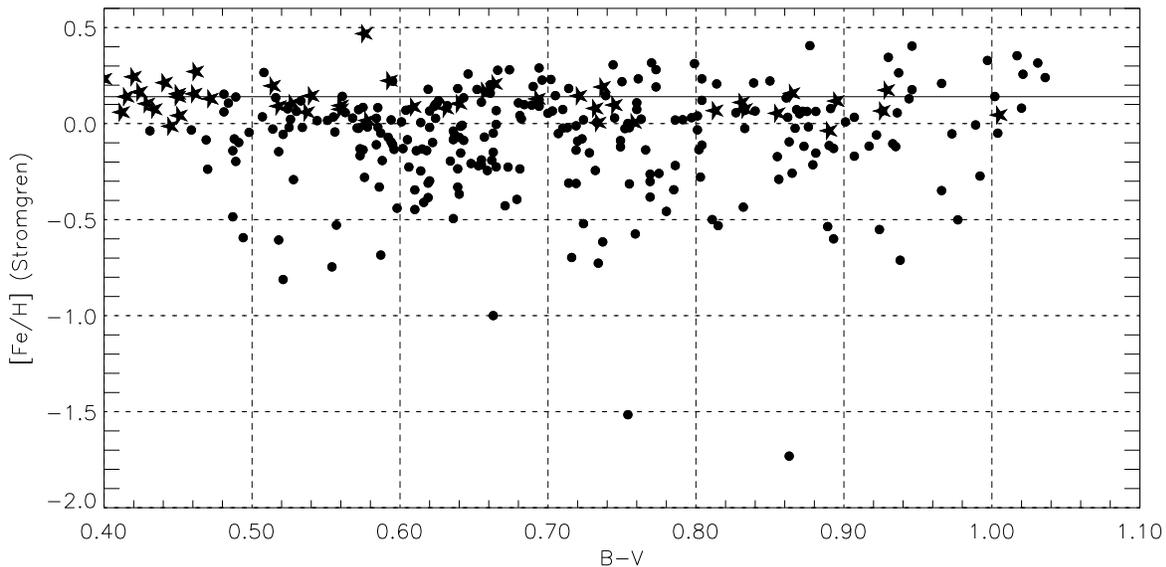}
\caption{
Corrected metallicity distribution as a function of 
B-V for the Str\"omgren photometry sample.
The horizontal line represents the median metallicity of the Hyades cluster determined from a sample of spectroscopic metallicity
from Perryman et al. (1998), [Fe/H]=+0.14.
The filled star symbols  are Hyades members from de Bruijne et al. (2001).
All metallicities were calculated using the corrected calibration of SN as given by equations (2, 3 \& 4).
}
\end{figure*}

Fig. 8 shows the photometric metallicities calculated for a set of spectroscopic
standards.

\subsection{A calibration with reference to the Hyades sequence}

Due to its rather complicated form, the application of the calibration given above is somewhat
tedious. Also, the non-explicit dependence on the Hyades 
sequence makes it  difficult to understand the origin of problems such as the
one presented here, afterwards.
On the contrary, a calibration directly based on the Hyades sequence, such as the one
proposed by Olsen (1984) for G and K dwarfs, has a simple form, and can be straighforwardly improved 
as the Hyades sequence gets better defined and as more spectroscopic metallicities become available.
We used a Hyades sequence defined by the coefficients from Tab. 1.
These coefficients were derived by fitting the $m_1(b-y)$ 'Olsen' sequence to the Hyades data 
presented on Fig. 5.

Using {\it PIKAIA} with the same fitness function and calibrating stars as in the previous section,
we derive new calibrations as a function of $\delta m_1$ and  $\delta c_1$.\\
We find the following relation valid for $0.22<b-y<0.37$ :

\begin{eqnarray}
[Fe/H]_{uvby}=0.108-14.91 \delta m_1    
\end{eqnarray}

and then 

\begin{eqnarray}
[Fe/H]_{uvby}=0.0783-9.095 \delta m_1 - 8.575(b-y)\delta c_1
\end{eqnarray}\\
for $0.37<b-y<0.59$.\\

These calibrations are illustrated on Fig. 9.
\begin{table}
 \centering
 \begin{minipage}{80mm}
  \caption{Coefficients of the polynomials used to fit the observed Hyades sequence, and utilised
to calculate the  $\delta m_1$ and  $\delta c_1$ indices (sec. 3.5). Second column gives the coefficients to calculate
$m_1$ as a function of $b-y$, third gives the coefficients to calculate $c_1$ as a function of $b-y$.
The fits are valid between $0.07<b-y<0.8$.
}
  \begin{tabular}{@{}llr@{}}    \hline
0 &  -0.109 &   +1.09\\   
1 &  +10.03 &   +1.85\\   
2 & -123.43 &  -56.76\\   
3 & +786.09 &  +390.12\\   
4 & -2927.5 & -1747.9\\   
5 &+6514.42 & +5031.2\\   
6 &-8388.02 & -8424.2\\   
7 &+5738.56 & +7347.46\\   
8 & -1613.7 & -2570.77\\    \hline
 \end{tabular} 
\label{distrib}
\end{minipage}
\end{table}

\section{Some consequences and conclusion}

Using the new calibration, it is now possible to recalculate  metallicities for the set of local dwarfs. 
The new (B-V, [Fe/H]) distribution is shown on Fig. 10. The result is now satisfactory, but the occasion is taken
to emphasise the need for a larger number of spectroscopic metallicities for cool dwarfs.

In the last decade, the calibration of SN have been widely used in various studies, in particular 
to design metallicity distribution of long-lived dwarfs in the solar neighbourhood.
In one such study, we  sampled the solar neighbourhood within 20~pc (Haywood, 2001).
The sample, before the selection of long-lived dwarfs, contained 177 stars for which metallicity
came from Geneva photometry, and 172 stars for which metallicity was calculated from Str\"omgren metallicity,
mostly for stars bluer than B-V=0.67.  Str\"omgren photometry was also used for 41 stars redder than this limit, for which
no Geneva photometry was available. All stars with SN metallicities were corrected as  [Fe/H]$_{SN}$/0.865+0.052.
Only 20 stars with colour in the critical interval 0.8$<B-V<$1.0 were included in the sample.
In view of the limited number of objects affected 
by the problem, it is unlikely that our metallicity distribution has suffered much from this effect. 
We conclude that the trend seen on  Fig. 1a between  metallicity  and colour is obviously real.

As a last check on our metallicities, we have searched for spectroscopic iron abundances for the stars in Fig. 1a in the 
catalogue of Cayrel et al. (2001). Fig. 11  shows both the photometric and spectroscopic
values $vs$ B-V colour and there is  good general agreement between the two. 
The spectroscopic metallicities, although still very sparse, confirm the general shape of the B-V-[Fe/H] distribution, and we conclude that our
findings of Haywood (2001) are robust.

\begin{figure*}
\includegraphics[width=16cm]{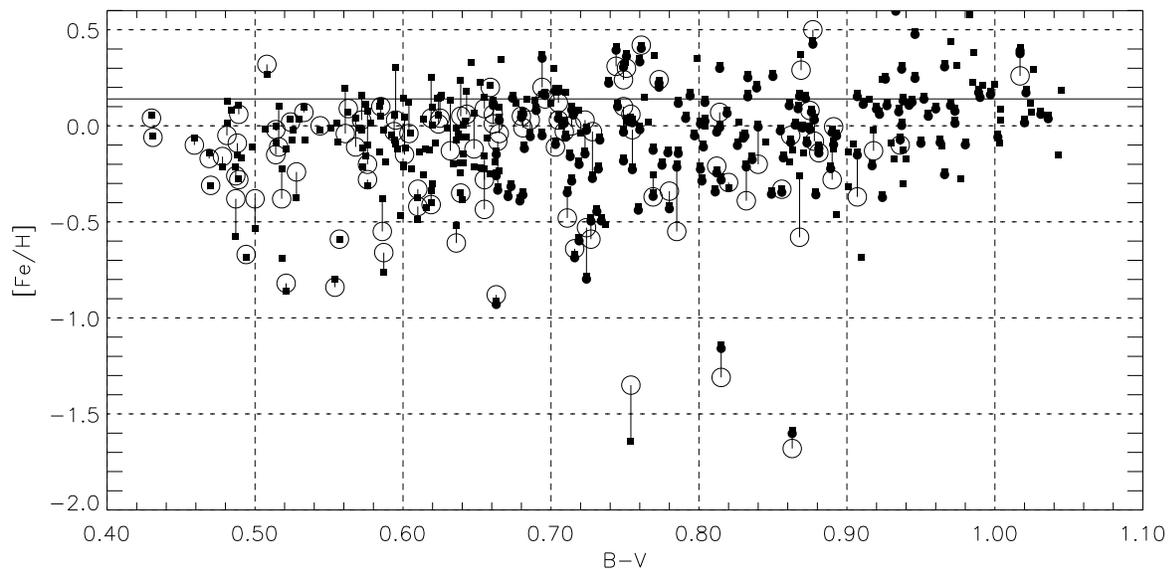}
\caption{B-V-[Fe/H] distribution for the sample in Haywood (2001).  Circles represent stars for which spectroscopic
measurements are available from the catalogue of Cayrel et al. (2001). The photometric measurements
are shown as squares, and related to the spectroscopic value with a vertical line. 
This plot shows that there is no systematic deviation between spectroscopic metallicities 
and our photometric abundances.
}
\end{figure*}

\section*{Acknowledgments}    
This research has made use of the SIMBAD database,
operated at CDS, Strasbourg, France, 
and of The General Catalogue of Photometric Data, University of Lausanne, Switzerland.
We thank the referee and Frederic Arenou for their suggestions and comments.

\end{document}